\documentclass[reprint,amsmath,amssymb,aps,prl,longbibliography]{revtex4-1}

\usepackage{graphicx}
\usepackage{dcolumn}
\usepackage{bm}
\usepackage{xcolor}
\usepackage{mathptmx}
\usepackage{comment}
\usepackage{physics}

\begin{document}

\title{Chebyshev self-imaging from inverse-sampled angular spectra}

\author{Layton A. Hall}
\email{laytonh@lanl.gov}
\affiliation{Los Alamos National Laboratory, Los Alamos, NM 87545, USA}
\author{Murat Yessenov}
\affiliation{Harvard John A. Paulson School of Engineering and Applied Sciences, Harvard University, Cambridge, MA, USA}

\begin{abstract}
For nearly two centuries, conventional self-imaging has obeyed a single rule: a periodic wave reconstructs itself at a fixed distance, independent of its bandwidth. Here we show this law is programmable. Sampling the angular spectrum in inverse powers of the mode index makes each mode rephase with its own integer period, deferring exact revival to the least common multiple of all periods: $z_{\mathrm{rev}}=z_{\mathrm{T}}\,e^{r\psi(M)}$, with $\psi$ the second Chebyshev function. We observe these revivals optically as a prime-power staircase in propagation distance, turning free-space diffraction into a readout of the multiplicative structure of the integers.
\end{abstract}

\maketitle

\emph{Introduction ---} Self-imaging is among the most robust collective effects in wave physics where a transversely periodic field reconstructs itself without lenses at integer multiples of the Talbot length $z_{\mathrm{T}}$ \cite{Talbot36PM, Rayleigh81PM}. At its root, self-imaging is a statement about phase commensurability. Under paraxial propagation, each spatial-frequency component $k_x$ acquires a quadratic phase $\propto k_x^2 z$, and for a periodic field the harmonics $k_x \propto m$ carry phases $\propto m^2$, a set of integers whose simultaneous return to unity fixes a single, bandwidth-independent revival distance. The same quadratic arithmetic dictates self-imaging across platforms including the temporal Talbot effect \cite{Jannson81JOSA, Andrekson93OL}, in waveguide arrays and photonic lattices \cite{Iwanow05PRL}, in matter waves and Bose--Einstein condensates \cite{Chapman95PRA,Deng99PRL}, spatio-temporal structured fields \cite{Yessenov20PRL1,Hall21arxivTalbot,HallOL21}, canonical embeddings \cite{HallOptica26}, and in the fractional images and quantum carpets that have become a signature of coherent revivals \cite{Berry96JMO,Berry01PW}.

Because the integers enter so directly into the self-imaging phase, it is natural to ask whether wave recurrences can be made to expose, or even compute, number-theoretic structure. The quadratic Talbot phase is precisely the kernel of a Gauss sum, and a substantial body of work has exploited optical and matter-wave interferometry to evaluate Gauss sums and, through them, to factor integers \cite{Bigourd08PRL,Mehring07PRL,Mack02PSSB,Merkel11NJP,Pelka18OE}. In every such scheme, the propagation arithmetic stays quadratic, inherited from the fixed harmonic spectrum of a grating; the number theory enters through how a chosen integer is read against that quadratic kernel, not through the self-imaging law itself.

Here we show that engineering the angular spectrum changes the arithmetic of self-imaging at its root. In conventional self-imaging, every mode rephases an integer number of times within one Talbot period --- all rephasing clocks are commensurate with the slowest --- so the field revives at a single universal distance regardless of bandwidth. Sampling the spectrum in inverse powers of the mode index, $k_x(m) = k_{\mathrm{L}}m^{-r/2}$, inverts this arithmetic: mode $m$ now rephases with its own integer period $\propto m^{r}$, and the field revives only where all these periods coincide, at their least common multiple. The first exact revival thus occurs at $z_{\mathrm{rev}} = z_{\mathrm{T}}\, \operatorname{lcm}(1,\dots,M)^{r} = z_{\mathrm{T}}\, e^{r\psi(M)}$, with $\psi$ the second Chebyshev function of number theory \cite{Apostol_AnalyticNumberTheory}. The recurrence thereby inherits the multiplicative rather than additive structure of the integers: the revival distance is constant as modes are added and jumps by exactly $p^{r}$ only when the mode index reaches a new prime power $p^{a}$. Self-imaging thus acquires a prime-power staircase, and the finite spectral cutoff $M$ is no longer a truncation but the quantity that fixes the exact recurrence.

\emph{Theoretical treatment ---} Self-imaging can be viewed as a rephasing of the spatial spectrum. Consider a scalar paraxial field as a finite superposition of $M$ transverse plane waves,
\begin{equation}
E(x,z)=\sum_{m=1}^{M}a_m\,e^{ik_x(m)x}\,e^{-ik_x^2(m)z/2k_\mathrm{o}},
\label{eq:field}
\end{equation}
where $k_\mathrm{o}$ is the optical wavenumber and $a_m$ the complex amplitude of the mode with transverse spatial frequency $k_x(m)$. The dynamics are set entirely by the quadratic phase $k_x^2 z/2k_\mathrm{o}$, so engineering the sequence $k_x^2(m)$ engineers the arithmetic of the modal rephasing.

\begin{figure*}[t!]
\centering
\includegraphics[width=16 cm]{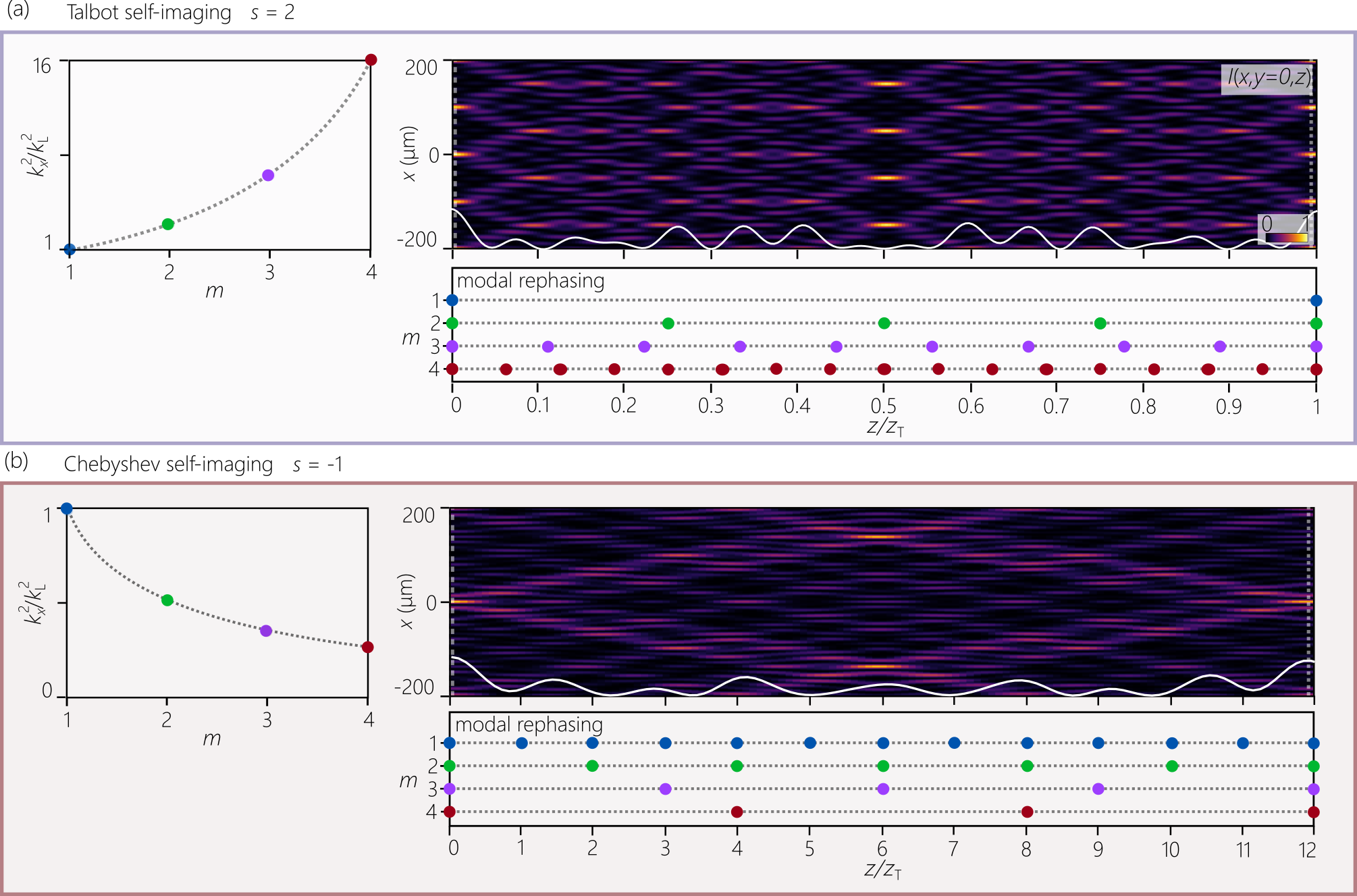}
\caption{Additive versus multiplicative self-imaging for power-law spectra $k_x(m)=k_{\mathrm{L}}m^{s/2}$, illustrated for $M=4$: (a)~Talbot ($s=2$) and (b)~Chebyshev ($s=-1$) self-imaging. \emph{Left:} sampled spectrum $k_x^2/k_{\mathrm{L}}^2=m^{s}$ with occupied modes $m=1$--$4$. \emph{Right:} simulated intensity carpet $I(x,y\!=\!0,z)$ with on-axis intensity (white) and, below, the modal rephasing planes $\zeta\in m^{-s}\mathbb{Z}$ at which each mode returns to its launch phase; self-imaging occurs where all four rows align. For $s=2$ (a) all periods $1/m^{s}\le1$ nest within the unit interval, so the field revives at $\zeta=1$ independently of $M$; for $s=-1$ (b) the periods are the integers $1$--$4$, deferring the first exact self-image to $\zeta=\mathrm{lcm}(1,2,3,4)=12$.}
\label{Fig:theory}
\end{figure*}

We choose a power-law angular spectrum $k_x(m)=k_{\mathrm{L}}m^{s/2}$ with $s$ being an integer. Defining the dimensionless coordinate $\zeta=z/z_{\mathrm{T}}$ with $z_{\mathrm{T}}=4\pi k_\mathrm{o}/k_{\mathrm{L}}^2$, each modal phase becomes $\exp(-2\pi i\,\zeta m^s)$. Strict field self-imaging, $E(x,z)=E(x,0)$, then requires $\zeta m^s\in\mathbb{Z}$ for all $1\le m\le M$.
This geometric picture, illustrated in Fig.~\ref{Fig:theory}, makes the qualitative dependence on the sign of $s$ immediate. We note that $s=1$ corresponds to the Montgomery effect~\cite{Montgomery67JOSA,Yessenov26Optica} and $s=2$ to the Talbot effect~\cite{Talbot36PM}.


For $s>0$ the $m^{s}$ are integers, so every rephasing period $m^{-s}\le1$ divides the unit interval: the modal lattices nest, all rows align at every integer plane [Fig.~\ref{Fig:theory}(a)], and the field revives at $\zeta_{\mathrm{rev}}=1$ independently of $M$, recovering conventional additive (Talbot-type) commensurability. Adding modes reshapes the carpet but never the revival distance.

For $s=-r$, the phases are $\exp(-2\pi i\,\zeta/m^r)$ and the arithmetic inverts, and the rephasing periods become the integers $m^r$ themselves, growing with mode index rather than nesting within the unit interval [Fig.~\ref{Fig:theory}(b)]. The self-imaging condition now demands $\zeta$ be a common multiple of $\{m^r\}_{m=1}^{M}$, so the first revival is
\begin{equation}
z_{\mathrm{rev}}(M)=z_{\mathrm{T}}\,\operatorname{lcm}(1,2,\ldots,M)^r,
\label{eq:revival}
\end{equation}
using $\operatorname{lcm}(1^r,\ldots,M^r)=\operatorname{lcm}(1,\ldots,M)^r$; in Fig.~\ref{Fig:theory}(b) the four rows first realign at $\zeta=\mathrm{lcm}(1,\ldots,4)=12$, and the carpet remains unrevived at every intermediate integer plane. Writing $\Lambda(M)=\operatorname{lcm}(1,\ldots,M)=\prod_{p\le M}p^{\lfloor\log_p M\rfloor}$, this revival coordinate $L_r(M)=\Lambda(M)^r$ obeys $\log\Lambda(M)=\psi(M)$, with $\psi$ the second Chebyshev function \cite{Nair1982,Chebyshev1852}, so
\begin{equation}
\frac{z_{\mathrm{rev}}(M)}{z_{\mathrm{T}}} = L_r(M) = e^{r\psi(M)}.
\label{eq:chebyshev}
\end{equation}
The exact recurrence is thus governed by the prime-power content of the occupied bandwidth. Hence, we call these \emph{Chebyshev revivals}.

The optical signature is a prime-power staircase where $\Lambda(M)$ changes only when $M$ reaches a new prime power,
\begin{equation}
\frac{L_r(M)}{L_r(M-1)}=
\begin{cases}
p^r, & M=p^a,\ p\ \text{prime},\\
1, & \text{otherwise}.
\end{cases}
\label{eq:jumps}
\end{equation}
Adding modes thus changes the revival distance only at prime-power indices. For $r=1$, $L_1(M)=12,60,60,420,840,2520$ at $M=4\text{--}9$, the jumps at $5$, $7$, $8=2^3$, and $9=3^2$ being direct fingerprints of the spectral prime-power structure.

\begin{figure*}[t!]
\centering
\includegraphics[width=\textwidth]{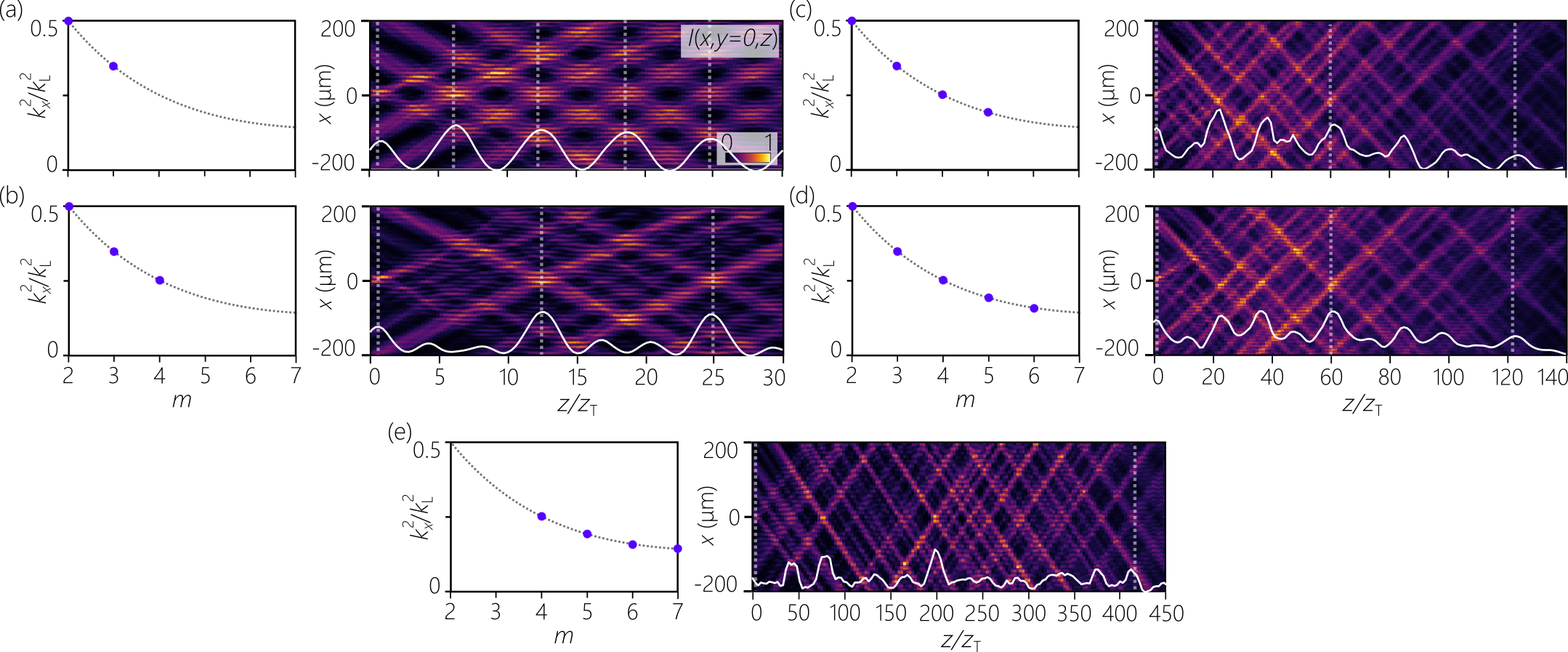}
\caption{Observation of Chebyshev revivals for inverse-sampled spectra $k_x(m)=k_{\mathrm{L}}m^{-1/2}$ ($r=1$). Each panel: sampled spectrum with occupied modes $m\in\mathcal{S}$ (left) and measured axial intensity $I(x,y\!=\!0,z)$ (right); white curves show the on-axis intensity $F(\zeta)$, vertical dashed lines the predicted revival planes at integer multiples of $\mathrm{lcm}(\mathcal{S})$. (a--d)~Consecutive spectra $\mathcal{S}=\{2,\dots,M\}$ for $M=3$--$6$, with first revivals at $\zeta=6$, $12$, $60$, and $60$ ($z_{\mathrm{T}}=0.5$, $0.5$, $0.15$, and $0.15$~mm, respectively): the revival distance is unchanged from $M=5$ to $6$, the plateau between prime powers. (e)~Sparse set $\mathcal{S}=\{4,5,6,7\}$ ($z_{\mathrm{T}}=0.03$~mm), first revival at $\zeta=\mathrm{lcm}(4,5,6,7)=420=\mathrm{lcm}(1,\dots,7)$.}
\label{Fig:fullset}
\end{figure*}

\begin{figure*}[t!]
\centering
\includegraphics[width=\textwidth]{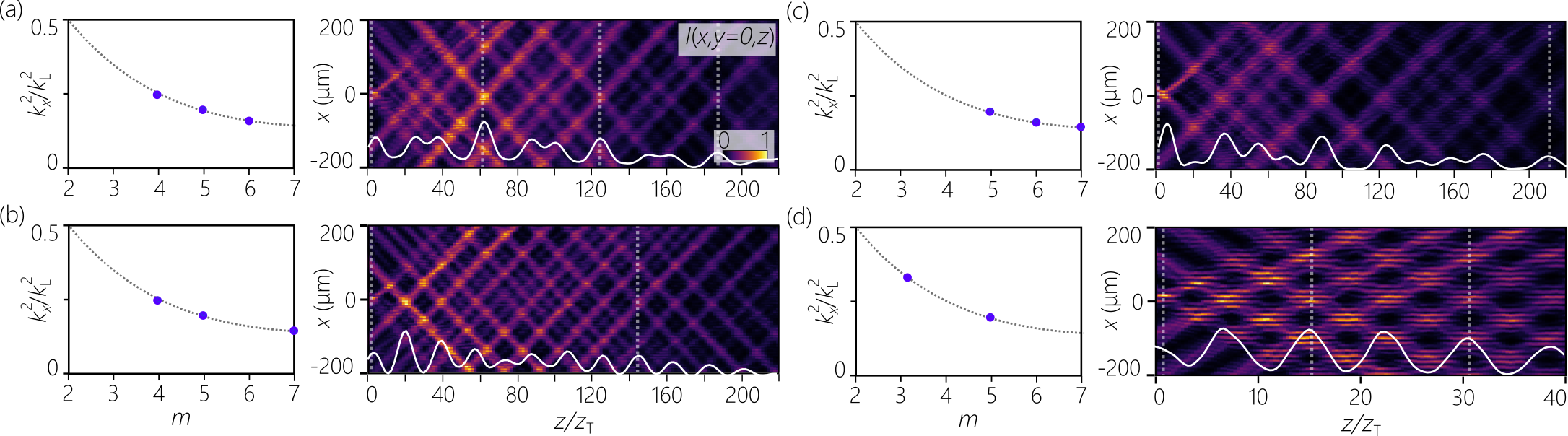}
\caption{Chebyshev revivals for sparse (non-consecutive) mode sets $\mathcal{S}$, for which the first revival occurs at $\zeta=\mathrm{lcm}(\mathcal{S})$ ($z_{\mathrm{T}}=0.1$~mm for (a--c) and $z_{\mathrm{T}} = 0.5$~mm for (d)). (a)~$\mathcal{S}=\{4,5,6\}$, $\mathrm{lcm}=60$; (b)~$\mathcal{S}=\{4,5,7\}$, $\mathrm{lcm}=140$; (c)~$\mathcal{S}=\{5,6,7\}$, $\mathrm{lcm}=210$; (d)~$\mathcal{S}=\{3,5\}$, $\mathrm{lcm}=15$.}
\label{Fig:subset}
\end{figure*}

The approach to the Chebyshev revival is itself structured by arithmetic, as the rephasing diagram of Fig.~\ref{Fig:theory}(b) already reveals. At intermediate integer planes, \emph{subsets} of the rows align. For $r=1$, mode $m$ accumulates phase $-2\pi\zeta/m$ and thus realigns at every plane $\zeta$ that is a multiple of $m$. At each integer plane, the modes whose index divides $\zeta$ therefore form an exact self-image of that part of the spectrum, while the remaining modes carry residual phases $e^{-2\pi i\zeta/m}$. We term these partial recurrences \emph{divisor revivals} [e.g., in Fig.~\ref{Fig:theory}(b), $\zeta=6$ aligns $m=1,2,3$ but not $m=4$]. Their hierarchy appears in the on-axis intensity $F(\zeta) = I(0,\zeta)/I(0,0)$, in Figs.~\ref{Fig:fullset} and~\ref{Fig:subset}. Partial peaks occur at every integer plane, tallest where most mode indices divide $\zeta$ [e.g., $\zeta=12$ for $M=5$, where only $m=5$ is misaligned], and $F=1$ is reached first at $\zeta=\mathrm{lcm}(1,\dots,M)$. The Chebyshev plane is thus the first \emph{exact} recurrence, not merely the strongest near-revival. This contrasts with the fractional Talbot effect, where intermediate planes carry Gauss-sum--weighted copies of the full field; here they instead select divisors from the spectrum, so the carpet between launch and revival renders the divisor lattice of the integers up to $M$.

\begin{figure}[t!]
\centering
\includegraphics[width=8.6 cm]{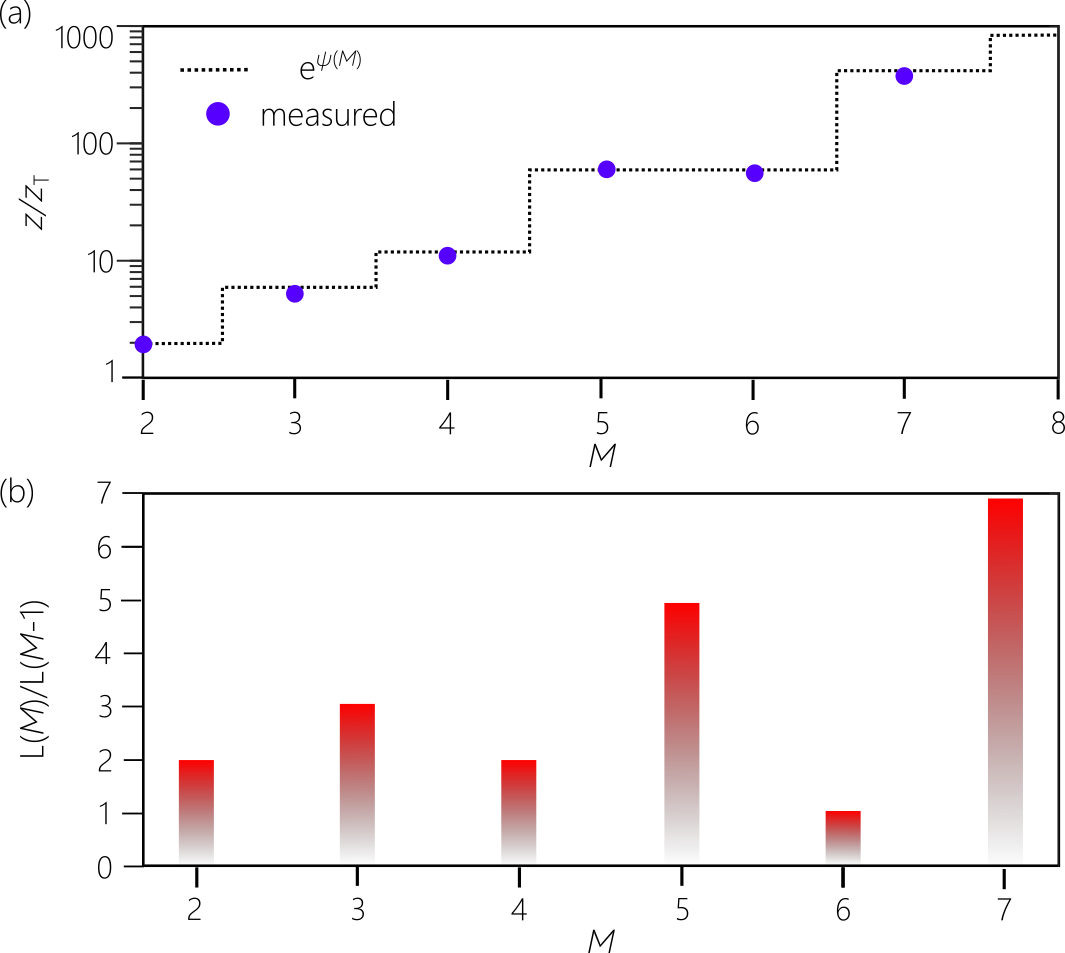}
\caption{The prime-power staircase of self-imaging from Fig. \ref{Fig:fullset}. (a)~Measured first-revival distance $z_{\mathrm{rev}}/z_{\mathrm{T}}$ versus occupied bandwidth $M$ (points), compared with the prediction $e^{\psi(M)}=\mathrm{lcm}(1,\dots,M)$ (dotted staircase), with $\psi$ the second Chebyshev function [Eq.~\eqref{eq:chebyshev}]. (b)~Measured ratio $L_1(M)/L_1(M-1)$ of consecutive revival distances.}
\label{Fig:staircase}
\end{figure}

\emph{Experimental demonstration ---} A CW laser ($\lambda=635$~nm) illuminates a phase-only SLM on which the occupied modes $m\in\mathcal{S}$, and hence the arithmetic of the revival, are programmed directly into a single row-interleaved hologram imparting $\Phi(x,y_m)=k_{\mathrm{L}}|x|/\sqrt{m}$ for $r=1$; the axial intensity $I(x,y\!=\!0,z)$ is recorded with a translated camera (setup and encoding details in the Supplemental Material).

We first prepare consecutive spectra \(\mathcal{S} = \{2,\dots,M\}\) for \(M = 3\)--\(6\). We omit the $m=1$ mode, the largest $k_x$, to relax the aperture requirement without altering the arithmetic, since $\mathrm{lcm}(2,\ldots,M)=\mathrm{lcm}(1,\ldots,M)$. The measured axial intensity carpets \(I(x,y\!=\!0,z)\) are shown in Fig.~\ref{Fig:fullset}, together with the on-axis intensity \(F(\zeta)\). For \(M=3\) and \(M=4\) [Figs.~\ref{Fig:fullset}(a,b)] the field revives at integer multiples of \(\zeta = 6\) and \(12\), respectively, in agreement with Eq.~\eqref{eq:revival}. Between revivals, \(F(\zeta)\) displays the predicted hierarchy of divisor revivals: partial peaks at every integer plane, with height set by the weight of the divisor-selected subset \(\{m : m \,|\, \zeta\}\). Increasing the bandwidth to \(M=5\) [Fig.~\ref{Fig:fullset}(c)] stretches the first full recurrence fivefold, to \(\zeta = \mathrm{lcm}(1,\dots,5) = 60\), marking the arrival of the prime \(5\) in the spectrum. Crucially, adding the \emph{sixth} mode [Fig.~\ref{Fig:fullset}(d)] leaves the revival distance pinned at \(\zeta = 60\), since \(6 = 2\cdot 3\) contributes no new prime power. The transverse pattern is reorganized, but the recurrence is not. This plateau is the qualitative departure from all quadratic (Talbot-type) self-imaging, in which the revival distance is insensitive to the occupied bandwidth altogether. Finally, the four-mode subset \(\mathcal{S}=\{4,5,6,7\}\) [Fig.~\ref{Fig:fullset}(e); \(z_{\mathrm{T}}=0.03\)~mm] extends the measured recurrences by a further factor of seven, to \(\zeta=\mathrm{lcm}(4,5,6,7)=420\). Because these four modes already exhaust the prime-power content of \(\{1,\dots,7\}\), we have \(\mathrm{lcm}(4,5,6,7)=\mathrm{lcm}(1,\dots,7)\), and the full \(M=7\) revival distance is attained with only four modes.

The full staircase is assembled in Fig.~\ref{Fig:staircase}(a), where we extract the first-revival distance from the intensity traces for \(M = 2\)--\(6\), together with the sparse set of Fig.~\ref{Fig:fullset}(e), which attains the full \(M=7\) distance, and compare it with \(e^{\psi(M)}\). The measured points track the exponentiated Chebyshev function over more than two decades in propagation distance.
The prime-power fingerprint is sharpest in the ratio of consecutive
revival distances, Fig.~\ref{Fig:staircase}(b): $L_1(M)/L_1(M-1)$
takes the values $2,3,2,5$ at $M=2,3,4,5$ --- the step of $2$ at
$M=4$ marking the prime power $2^2$ --- collapses to unity at the
composite $M=6$, and equals $7$ at the prime $M=7$, realizing
Eq.~\eqref{eq:jumps} directly.
The propagation distance thus performs a primality (more precisely, a prime-power) test on the mode index. Whether \(M\) is a prime power is read off directly from whether the carpet lengthens. Measurements for \(r=2\), which square every revival coordinate and realize the \(p^{r}\) jumps of Eq.~\eqref{eq:jumps} directly, are presented in the End Matter.

Because the revival coordinate is \(\mathrm{lcm}(\mathcal{S})\) for an \emph{arbitrary} mode set, the recurrence is programmable beyond consecutive bandwidths. In Fig.~\ref{Fig:subset} we prepare the sparse spectra \(\mathcal{S}=\{4,5,6\}\), \(\{4,5,7\}\), \(\{5,6,7\}\), and \(\{3,5\}\), with first revivals at \(\zeta=60\), \(140\), \(210\), and \(15\). Removing low-order modes coarsens the divisor lattice, so fewer partial revivals punctuate the carpet; two spectra of equal cardinality can differ in revival distance by an order of magnitude, and a two-mode field [Fig.~\ref{Fig:subset}(d)] can be endowed with any revival \(\mathrm{lcm}(m_1,m_2)\). In (d) the \emph{intensity} additionally recurs at \(\zeta=15/2\), where both modes acquire a common phase \(\pi\) and the field revives only up to a global sign; the first strict field revival remains at \(\zeta=15\).

\emph{Discussion ---} These results establish that the revival distance of a wave is not a fixed property of propagation but a programmable arithmetic quantity: by shaping only the modal phase rates, we set the recurrence by the multiplicative structure of the integers — which primes and prime powers the spectrum contains — rather than by the additive commensurability that governs Talbot imaging. The distinction from Gauss-sum
factorization schemes \cite{Bigourd08PRL,Mehring07PRL,Mack02PSSB, Merkel11NJP,Pelka18OE} is essential. In those schemes, arithmetic is encoded in a readout performed against a fixed quadratic propagation kernel, whereas here the arithmetic resides in the propagation law itself. The carpet between launch and revival is a physical rendering of the divisor lattice, and the first exact recurrence computes $\operatorname{lcm}$ and, through it, $\psi(M)$.

The inverse-sampled spectrum has a natural quantum counterpart in the Coulomb problem, whose bound-state energies $E_n\propto -1/n^2$ realize precisely the $r=2$ phase law in the time domain (see End Matter for experimental plots). The literature on Rydberg wave-packet revivals \cite{Parker86PRL,Averbukh89PLA,Yeazell91PRA, Robinett04PR} treats superpositions centered on a large mean quantum number $\bar{n}$, where Taylor expansion of the spectrum yields a hierarchy of approximate revivals and superrevivals. The Chebyshev revivals reported here are the exact-arithmetic limit of that physics. For a few-mode superposition, no expansion is required, and the recurrence is fixed rigorously by the $\operatorname{lcm}$ of the occupied indices. Our optical construction thus serves as a programmable classical emulator of Coulomb-like revival dynamics and, conversely, suggests that engineered few-level Rydberg superpositions prepared by mode-selective excitation should exhibit the same prime-power staircase in time, with the divisor revivals furnishing a spectroscopic divisibility test on the populated levels.

The prime-number theorem, $\psi(M)\sim M$, implies that the revival distance grows exponentially with bandwidth, $z_{\mathrm{rev}}\sim z_{\mathrm{T}}\,e^{rM}$; in free space the accessible $M$ is thus set by the interrogable propagation length, the transverse aperture, and phase-precision requirements that tighten exponentially with $M$ (see Supplemental Material) --- the revival encodes $\operatorname{lcm}(1,\ldots,M)$ as an exact physical recurrence, not a scalable computation of it. Because the construction relies only on programming the modal phase rates, Chebyshev revivals port to every platform hosting the Talbot effect \cite{Jannson81JOSA,Azana01JSTQE,Iwanow05PRL,Chapman95PRA}, and dispersive fiber offers the most direct route past this bound: a pulse shaper assigning frequencies $\omega(m)$ with group-delay-dispersion phase $\beta_{2}\,\omega^{2}(m)\,z/2 \propto z/m^{r}$ maps $z_{\mathrm{T}}$ onto meters of fiber, and kilometer-scale low-loss propagation \cite{Andrekson93OL} brings $\operatorname{lcm}(1,\ldots,M)\sim 10^{4}$, i.e.\ $M\approx 10$--$12$, within reach. More broadly, the finite aperture of a synthesized spectrum need not be a mere truncation but is instead a programmable arithmetic register, read out by free-space diffraction.

\noindent\textbf{Funding}
Los Alamos National Laboratory LDRD program grant 20251140PRD1. 

\noindent\textbf{Acknowledgments}
The authors thank M. Martin for the equipment. 

\noindent\textbf{Disclosures}
The authors declare no conflicts of interest.

\noindent\textbf{Data availability}
Data underlying the results presented in this paper are available upon reasonable request.

\appendix
\begin{center}
\textbf{End Matter}
\end{center}

\begin{figure}[h!]
\centering
\includegraphics[width=\columnwidth]{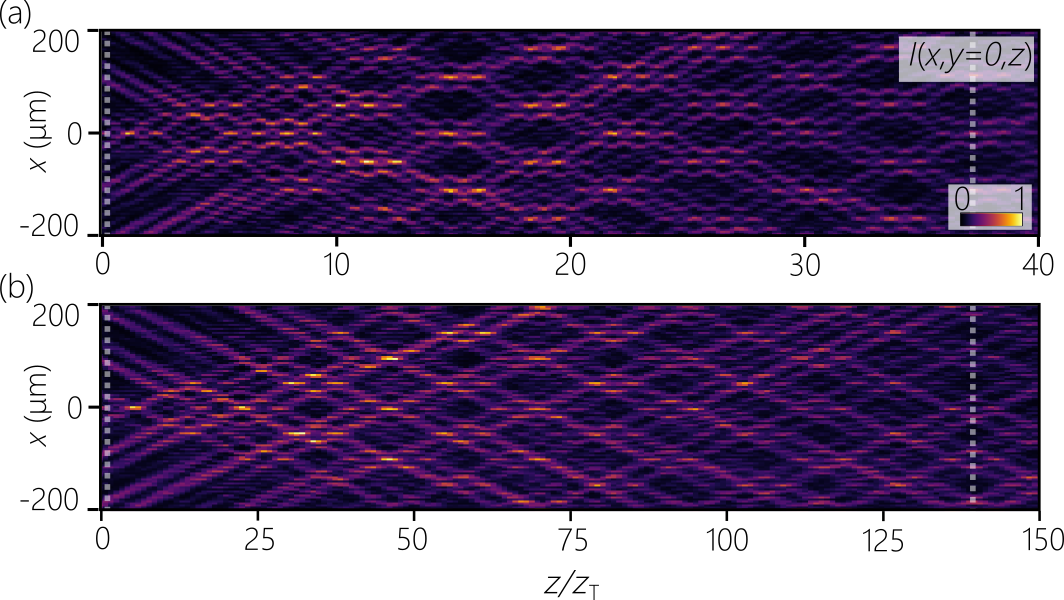}
\caption{Chebyshev revivals for $r=2$, $k_x(m)=k_{\mathrm{L}}m^{-1}$. Measured axial intensity $I(x,y\!=\!0,z)$ for (a)~$\mathcal{S}=\{1,2,3\}$, first revival at $\zeta=L_2(3)=36$ ($z_{\mathrm{T}}=0.5$~mm), and (b)~$\mathcal{S}=\{2,3,4\}$, first revival at $\zeta=L_2(4)=144$ ($z_{\mathrm{T}}=0.1$~mm).} 
\label{Fig:r2}
\end{figure}

\medskip
\noindent\textit{Chebyshev revivals for $r=2$} --- To isolate the role of the exponent $r$, we prepare spectra sampled in the inverse first power of the mode index, $k_x(m)=k_{\mathrm{L}}m^{-1}$ ($s=-2$), so that each modal phase advances as $-2\pi\zeta/m^{2}$ and mode $m$ rephases only after every $m^2$ Talbot lengths. The first self-image is accordingly deferred to $\zeta=L_2(M)=\mathrm{lcm}(1,\dots,M)^{2}$. Figure~\ref{Fig:r2} shows the measured intensity carpets for the consecutive spectra $\mathcal{S}=\{1,2,3\}$ and $\{2,3,4\}$: the field revives at $\zeta=36$ and $144$, respectively, and the measured steps between them --- a factor of $9=3^{2}$ from $M=2$ to $3$, and of $4=2^{2}$ from $M=3$ to $4$ --- realize the $p^{r}$ jumps of Eq.~\eqref{eq:jumps}: the exponent magnifies each prime-power jump from $p$ to $p^{2}$. The partial-recurrence hierarchy is likewise squared: at an intermediate plane, only the modes whose \emph{squared} index divides $\zeta$ rephase, so the carpet renders the square-divisor lattice rather than the divisor lattice of $r=1$ [e.g., at $\zeta=36$ for $M=4$, only $m=4$ is misaligned, since $16\nmid 36$]. These $-2\pi\zeta/m^{2}$ phases are the spatial analog of the Coulomb spectrum $E_n\propto-1/n^{2}$, making the $r=2$ measurements a direct classical emulation of the Rydberg-like revival dynamics discussed in the main text.

\bibliography{diffraction}

\end{document}